\documentclass[twocolumn,showpacs,floats,floatfix,superscriptaddress,aps,prl]{revtex4}
\usepackage{amsfonts,amssymb,amsmath}
\usepackage{color,calc}
\usepackage[dvips]{graphicx}
\usepackage{bm}
\usepackage[normalem]{ulem}

\def\be{ \begin{equation} }
\def\ee{ \end{equation} }
\def\bea{ \begin{eqnarray} }
\def\eea{ \end{eqnarray} }
\def\bse{ \begin{subequations} }
\def\ese{ \end{subequations} }

\def\i{\,\text{i}}
\def\e{\,\text{e}}
\def\i{i}
\def\e{e}

\def\to{\rightarrow}

\newcommand{\ket}[1]{\vert #1\rangle}
\def\d{\text{d}}

\def\U{\mathbf{U}}

\def\H{\mathbf{H}}
\def\c{\mathbf{c}}

\def\ket#1{| #1 \rangle}
\def\bra#1{\langle #1 |}

\def\to{\rightarrow}
\def\jmax{j_{\text{max}}}
\def\jmax{j_0}
\def\i{\text{i}}
\def\f{\text{f}}

\newcommand{\pryso}{$\text{Pr}^{3+}\text{:}\text{Y}_2\text{SiO}_5\:$}

\begin{document}

\author{Genko T. Genov}
\affiliation{Department of Physics, St. Kliment Ohridski University of Sofia, 5 James Bourchier blvd, 1164 Sofia, Bulgaria}
\affiliation{Institut f{\"u}r Angewandte Physik, Technische Universit{\"a}t Darmstadt, Hochschulstr. 6, 64289 Darmstadt, Germany}
\author{Daniel Schraft}
\affiliation{Institut f{\"u}r Angewandte Physik, Technische Universit{\"a}t Darmstadt, Hochschulstr. 6, 64289 Darmstadt, Germany}
\author{Thomas Halfmann}
\affiliation{Institut f{\"u}r Angewandte Physik, Technische Universit{\"a}t Darmstadt, Hochschulstr. 6, 64289 Darmstadt, Germany}
\author{Nikolay V. Vitanov}
\affiliation{Department of Physics, St. Kliment Ohridski University of Sofia, 5 James Bourchier blvd, 1164 Sofia, Bulgaria}

\title{Correction of Arbitrary Errors in Population Inversion of Quantum Systems by Universal Composite Pulses}

\date{\today}

\begin{abstract}
We introduce universal broadband composite pulse sequences for robust high-fidelity population inversion in two-state quantum systems, which compensate deviations in \emph{any} experimental parameter (e.g. pulse amplitude, pulse duration, detuning from resonance, Stark shifts, unwanted frequency chirp, etc.) and are applicable with \emph{any} pulse shape.
We demonstrate the efficiency and universality of these composite pulses by experimental data on rephasing of atomic coherences in a \pryso crystal.
\end{abstract}

\maketitle


Composite pulses (CPs) have been used for decades in nuclear magnetic resonance \cite{NMR} and, since recently,
 in quantum information processing \cite{Blatt,Wunderlich,Hill,Ivanov11PRA} and quantum optics \cite{QOptics,Torosov11PRA,Torosov11PRL} for highly accurate and robust qubit rotation.
Similar ideas have been developed in applied optics even earlier, as a tool to design polarization filters and achromatic polarization retarders \cite{Optics}.
The basic idea of CPs is to correct the imperfect interaction of a quantum system with a single pulse by using a sequence of pulses with suitably chosen relative phases.
The errors introduced by the constituent pulses are canceled by destructive interference up to a certain order.

A common feature of CPs developed so far is that they compensate experimental variations in a single parameter only (e.g. pulse duration, pulse amplitude, detuning), or simultaneous fluctuations in at most two parameters \cite{NMR}. Moreover, the optimal phases of CPs usually depend on their pulse shape.
These features limit the advantages of CPs for applications requiring very high fidelity (e.g. in quantum computation) or robustness with regard to variations in several experimental parameters.

In this Letter, we describe a general theoretical procedure to derive universal CPs for complete population inversion,
 which compensate deviations in \emph{any} experimental parameter and work with \emph{any} pulse shape.
The only assumptions made are those of a two-state system, coherent evolution and identical pulses in the CP sequence.
As a basic example of relevance to many applications in quantum physics,
we experimentally demonstrate the concept by rephasing of atomic coherences for coherent optical data storage in a \pryso crystal.

%

We consider a coherently driven two-state quantum system.
Its dynamics obeys the Schr\"{o}dinger equation, $\i \hbar\partial_t \mathbf{c}(t) = \H(t)\mathbf{c}(t)$,
where the vector $\c(t) = [c_1(t), c_2(t)]^T$ contains the probability amplitudes of the two states.
The Hamiltonian in the rotating-wave approximation reads
$\mathbf{H}(t) = (\hbar/2)\Omega(t)\e^{-\i\delta(t)} \ket{1}\bra{2} +\text{H.c.} $,
with
$\delta(t)=\int_{0}^{t}\Delta(t^{\prime})\d t^{\prime}$, where $\Delta=\omega_0-\omega$ is the detuning between the field frequency $\omega$ and the Bohr transition frequency $\omega_0$.
The Rabi frequency $\Omega(t) =-\mathbf{d}\cdot\mathbf{E}(t)/\hbar$ defines the coupling of the two states, induced by the electric field $\mathbf{E}(t)$ and the transition dipole moment $\mathbf{d}$.
In general, both $\Omega(t)$ and $\Delta(t)$ are time-dependent.

We assume that the CP duration is shorter than the decoherence times in the system.
Then, the evolution is described by the propagator $\U$, which connects the amplitudes at the initial and final times $t_{\i}$ and $t_{\f}$:
 $\c(t_{\f})=\U \c(t_{\i})$.
It is conveniently parameterized with the three real St\"{u}ckelberg variables $q$ ($0\leqq q \leqq 1$), $\alpha$ and $\beta$ as
\be \label{2stateU}
\U = \left[\begin{array}{cc} q\, \e^{\i\alpha}  & p\, \e^{\i\beta} \\  -p\, \e^{-\i\beta} & q\, \e^{-\i\alpha} \end{array} \right],
\ee
where $p=\sqrt{1-q^2}$.
If the system is initially in state $|1\rangle$, then $p^2$ is the transition probability to state $|2\rangle$, and $q^2$ is the probability for no transition.
A constant phase shift $\phi$ in the Rabi frequency, $\Omega(t)\to\Omega(t)\e^{\i\phi}$, is imprinted in the propagator $\U(\phi)$ as
 \cite{Torosov11PRA}
\be \label{2stateUf}
\U(\phi)
=\left[\begin{array}{cc} q\, \e^{\i\alpha}  & p\, \e^{\i(\beta+\phi)} \\  -p\, \e^{-\i(\beta+\phi)} & q\, \e^{-\i\alpha} \end{array} \right].
\ee
The propagator for a composite sequence of $n$ pulses, each with a phase $\phi_{k}$, reads
\begin{align}
\U^{(n)} &= \U (\phi_{n})\cdots\U(\phi_{2})\U(\phi_{1}),  \label{U_SU2Comp}
\end{align}
where the phases $\phi_{1},\dots,\phi_{n}$ are free control parameters.

Our objective is to transfer all population from state $\ket{1}$ to state $\ket{2}$ even when the properties of the driving pulse are \emph{unknown}.
This requires robustness to deviations in all pulse parameters,
i.e., pulse shape, Rabi frequency, duration, detuning from resonance, dynamic Stark shifts, residual frequency chirp, etc.
Mathematically, our goal is to maximize the transition probability $P^{(n)} = |U^{(n)}_{21}|^2$, i.e. to minimize the CP infidelity (the probability for no transition) $Q^{(n)}=|U^{(n)}_{11}|^2$ for any values of $q$, $\alpha$, and $\beta$.
We make no assumptions about the constituent pulses, i.e. how $q$, $\alpha$ and $\beta$ depend on the interaction parameters.
This justifies the term ``universal'' for these CPs because they will compensate imperfections in \emph{any} interaction parameter.
We only assume that the constituent pulses are identical and that we have control over their phases $\phi_{k}$.

\begin{table}[bt]
\caption{Phases of universal CPs with $n$ pulses (indicated by the number in the label of the CP).
We have $\jmax=0$ for $n=3$, $\jmax=2$ for $n=5$ to $11$, $\jmax=4$ for $n=13$ and $\jmax=8$ for $n=25$.
Each phase is defined modulo $2\pi$.
The excitation dynamics remain the same when we simultaneously change the sign of all phases or add a constant shift to all phases.
} 
\begin{tabular}{l l} 
\hline 
Pulse & Phases \\ 
\hline 
 U3 & $(0, 1,0)\pi/2$ \\
U5a & $(0, 5, 2,5,0)\pi/6$ \\
 U5b & $(0, 11, 2,11,0)\pi/6$ \\
U7a & $(0, 11, 10, 17,10,11,0)\pi/12$ \\
 U7b & $(0, 23, 10, 5,10,23,0)\pi/12$ \\
U9a & $(0, 0.635, 1.35, 0.553, 0.297,0.553,1.35,0.635,0)\pi$ \\
 U9b & $(0, 1.635, 1.35, 1.553, 0.297,1.553,1.35,1.635,0)\pi$ \\
U11a & $(0, 0.574, 0.085, 0.378, 0.553, 1.105,0.553,0.378,$\\ & $0.085,0.574,0)\pi$ \\
U11b & $(0, 1.574, 0.085, 1.378, 0.553, 0.105,0.553,1.378,$\\ & $0.085,1.574,0)\pi$ \\
U13a & $(0, 9, 42, 11, 8, 37, 2,37,8,11,42,9,0)\pi/24$ \\
U13b & $(0, 33, 42, 35, 8, 13, 2,13,8,35,42,33,0)\pi/24$ \\
U25a & $(0, 5, 2, 5, 0, 11, 4, 1, 4, 11, 2, 7, 4,7,2,11,4,1,4,11,$ \\ & $0,5,2,5,0)\pi/6$ \\
U25b & $(0, 11, 2, 11, 0, 5, 4, 7, 4, 5, 2, 1, 4,1,2,5,4,7,4,5,$ \\ & $0,11,2,11,0)\pi/6$ \\ 
\hline 
\end{tabular}
\label{table:coef} 
\end{table}

\begin{figure}[tb]
 \includegraphics[width=1\columnwidth]{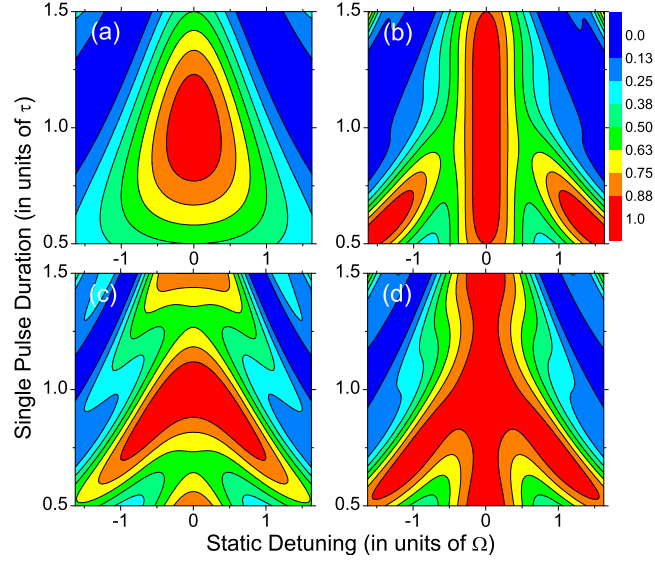}
\caption{(color online)
Numerically simulated fidelity $|U^{(n)}_{12}|^2$ vs. static detuning and duration $T$ of each constituent pulse (referred to as ``single pulse duration'') for
(a) a single rectangular pulse;
(b) three-pulse area-compensating CP with phases $(0,2\pi/3,0)$ \cite{Tycko84};
(c) three-pulse detuning-compensating CP with phases $(0,\pi/3,0)$ \cite{Torosov11PRA};
(d) universal CP with phases $(0,\pi/2,0)$ from Table \ref{table:coef}.
$\tau$ is the duration, for which the single-pulse area is equal to $\pi$, i.e. $\tau=\pi/\Omega$.
 }
\label{Fig-cp3}
\end{figure}

In order to determine the phases $\phi_{k}$ of a universal CP sequence of $n$ pulses we calculate the propagator element $U^{(n)}_{11}$ from Eqs.~\eqref{2stateUf} and \eqref{U_SU2Comp}.
We find
$U^{(n)}_{11}=\sum_{j=1}^{n} c_{nj} q^{j}$,
where the coefficients $c_{nj}$ depend on $\alpha$ and $\phi_{k}$ only.
Following arguments in Refs.~\cite{Torosov11PRA,Torosov11PRL} we use symmetric CPs, the phases of which obey the anagram relation $\phi_{n+1-k}=\phi_{k}$, with $k=1, 2,\dots, (n-1)/2$.
Such symmetric sequences automatically nullify all $c_{nj}$ for even $j$.
We choose phases such that the coefficients $c_{nj}$ are nullified for any $\alpha$ up to the highest possible order, denoted $\jmax$.
This generates a system of trigonometric equations for the phases $\phi_{k}$.
In general, there are multiple solutions. We choose solutions, which minimize the sum of the absolute values of the coefficients of the first non-zero order $\jmax+1$.
The transition probability for such CP reads
\be
P^{(n)} = 1-|U^{(n)}_{11}|^2 = 1-O(q^{2\jmax+2}).
\ee
Since $0\leqq q \leqq 1$, we can make the error term arbitrary small (unless $q=1$, which means zero transition probability for a single pulse)
 by nullifying $c_{nj}$ to an arbitrary high order $\jmax$ by using longer composite sequences.

For example, for a five-pulse sequence, we have
\begin{align}
U_{11}^{(5)} =& \{[1+2\cos(2\phi_2-\phi_3)] \e^{\i\alpha} + 2\cos(\phi_2-\phi_3) \e^{-\i\alpha}\} q \notag\\
 &+ O(q^3).
\end{align}
The $q$-term vanishes for two distinct sets: 
$\{\phi_2=5\pi/6$, $\phi_3=\pi/3\}$ and $\{\phi_2=11\pi/6$, $\phi_3=\pi/3\}$.
Then, we have $\jmax=2$ and the transition probability is $P^{(n)} = 1 - |U_{11}^{(5)}|^2 = 1- O(q^6)$.

\begin{figure*}
 \includegraphics[width=2\columnwidth]{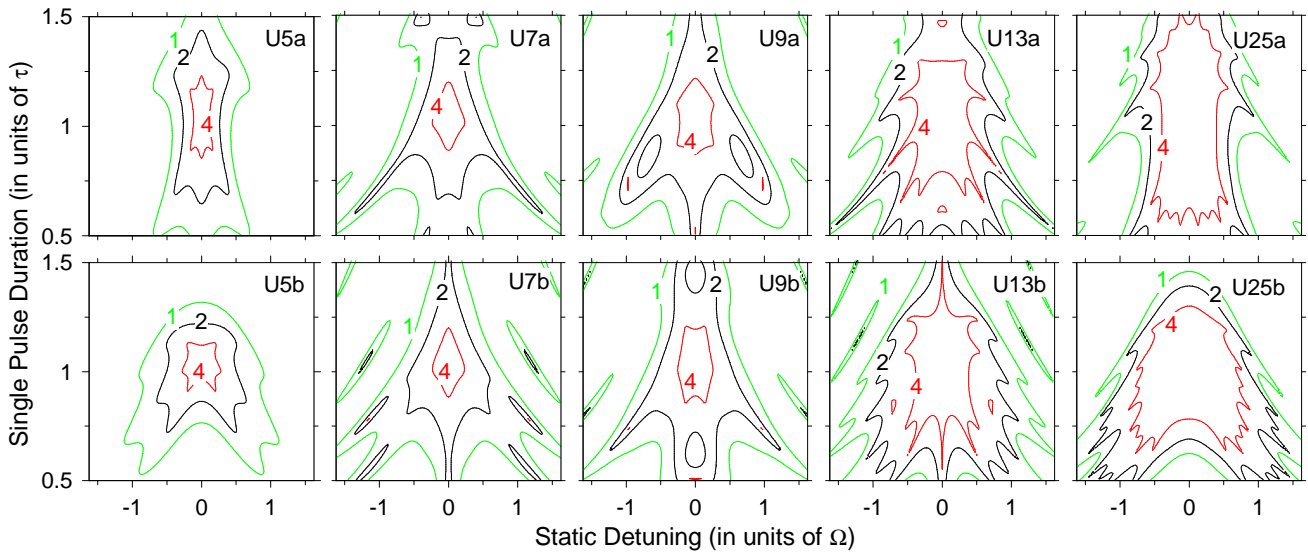}
\caption{(color online)
Numerically simulated infidelity $|U^{(n)}_{11}|^2$ vs. static detuning and single-pulse duration $T$ for several universal CPs (with rectangular shapes) from Table \ref{table:coef}.
The labels $m=1,2,4$ on the contour lines indicate the error level $10^{-m}$.
 }
\label{Fig-all}
\end{figure*}

Several universal CPs are listed in Table \ref{table:coef}.
The CPs labeled {``a''} perform better against variations in the pulse duration, whereas those labeled {``b''} perform better against the detuning.
Either of these, however, permit compensation against both parameters, as well as against \emph{all} other parameters.
Nearly all previously published CPs were derived for specific pulse properties, e.g. rectangular shape.
Such assumptions simplify the propagator,
 but when they are not met, the CPs may fail.
The CPs U5a and U5b in Table \ref{table:coef} were derived earlier, aiming at compensation of pulse area \cite{Tycko84}
and detuning errors \cite{Torosov11PRA} for specific pulse shapes.
We find now that these CPs indeed permit universal compensation of errors in any experimental parameter and for any pulse shape.
To the best of our knowledge, all other universal CPs in Table \ref{table:coef} were not known before at all.

The performance of the universal CPs versus deviations in the pulse duration $T$ of each constituent pulse and the detuning is shown in Figs.~\ref{Fig-cp3} and \ref{Fig-all}.
Figure \ref{Fig-cp3} compares a single pulse (frame (a)) to three-pulse CPs, which compensate the pulse duration alone (frame (b)), the detuning alone (frame (c)),
 and the universal CP U3 (frame (d)), which compensates both the pulse duration and the detuning.
As it is well known, the transition probability for a single pulse [Fig.~\ref{Fig-cp3}(a)] quickly drops when the pulse duration $T$ does not match to a perfect $\pi$-pulse, i.e., $T=\tau$, or when the pulse carrier frequency is off resonance.
The CP sequences in frames (b) and (c) of Fig.~\ref{Fig-cp3}, which compensate variations in a single parameter only, produce high-efficiency regions along one axis only.
The universal CP sequence U3 in Fig.~\ref{Fig-cp3} (d) is robust with respect to variations along both axes
 and its operation bandwidth is the largest one.

We note that the CP sequence U3 is not truly universal because the phases do not nullify the first-order coefficients (i.e., $\jmax=0$), but only minimize them, cf. Table~\ref{table:coef}.
Figure \ref{Fig-all} shows the performance of higher-order, genuine universal CPs. As expected, the high-fidelity region expands steadily, as the CP order increases.
We verified by extensive numerical simulations the robustness of the universal CPs against variations in other interaction parameters, e.g. Stark shifts, unwanted frequency chirp, pulse shape, frequency jitter, etc. All simulations confirm, that our universal CPs are amazingly robust to any such variation.
Note that in Fig.~\ref{Fig-all}, in contrast to Fig.~\ref{Fig-cp3}, we plotted the error (i.e. the infidelity) and we used a logarithmic scale in order to verify the suitability of these CPs for applications requiring ultra-high fidelity, e.g. as in quantum information processing.
Clearly, the low-error regions (with the label ``4'' indicating $10^{-4}$ error) are far broader than for a single pulse (a barely visible tiny spot for the latter), and they expand with the size of the CPs.


We experimentally verified the performance of the theoretically derived universal CP sequences by rephasing atomic coherences for optical data storage.
In the experiment, we generate the atomic coherence on a magnetic, radio-frequency (RF) transition between two inhomogeneously broadened hyperfine levels of a \pryso crystal.
The atomic coherence between the two quantum states is prepared  and read-out by electromagnetically-induced transparency (EIT) \cite{Fleischhauer05RMP}.
This enables straightforward optical readout.
The EIT scheme couples states $\vert 1\rangle$ and $\vert 2 \rangle$ by a strong control field and a weak probe field via a metastable excited state $\vert 3\rangle$.
By simultaneously and adiabatically turning off the control and probe fields, we convert the probe field into an atomic coherence, i.e. a coherent superposition of states $\vert 1\rangle$ and $\vert 2 \rangle$. This is the ``write'' process of optical information encoded in the probe field, often also termed ``stopped light'' or ``stored light'' \cite{Fleischhauer05RMP}.
To ``read'' the optical memory after an arbitrary storage time, we apply the strong control field again to beat with the atomic coherence and thereby generate a signal field with the same properties as the stored probe field. The concept and the experimental setup for EIT-based light storage in \pryso are described in detail elsewhere \cite{Schraft13PRA,Heinze13PRL,Mieth12PRA}.

In such a coherent optical memory it is crucial to reverse the effect of dephasing of the atomic coherences during the storage time. The dephasing in a doped solid such as \pryso is due to inhomogeneous broadenings of the hyperfine levels. Rephasing techniques serve to invert dephasing and provide large read-out signal also for longer storage times. In the simplest case, rephasing is implemented by resonant RF $\pi$-pulses, e.g. in a standard CPMG sequence \cite{CPMG}.
However, rephasing by resonant $\pi$-pulses does not work efficiently in systems with large inhomogeneous broadening, as the transition frequency varies for different ensembles within the inhomogeneous linewidth.
Hence, the driving pulse is no more a $\pi$-pulse for all ensembles.
The efficiency is further reduced by the spatial inhomogeneity of the RF field over the crystal.
As an alternative, adiabatic rephasing techniques, e.g. rapid adiabatic passage \cite{Pascual-Winter12PRB,Mieth12PRA} or composite adiabatic passage \cite{Torosov11PRL,Schraft13PRA} offer improved operation bandwidth with regard to variations in the parameters of the system and the driving pulses. To permit much broader operation bandwidth, we replace now the single $\pi$-pulses in the CPMG rephasing sequence by our universal CPs. In the experiment we set the storage time at $600\,\mu$s, which is much larger than the dephasing time $T_{\text{deph}} \approx 20\ \mu$s in $\text{Pr}^{3+}\text{:}\text{Y}_2\text{SiO}_5$. During the measurements the optical ``write'' and ``read'' sequences were kept fixed, while the RF rephasing pulses were varied.
Therefore, the energy of the retrieved signal serves as a straightforward measure of the rephasing efficiency, and hence, the efficiency of the driving $\pi$-pulse or CP sequence.

\begin{figure}
\includegraphics[width=1\columnwidth]{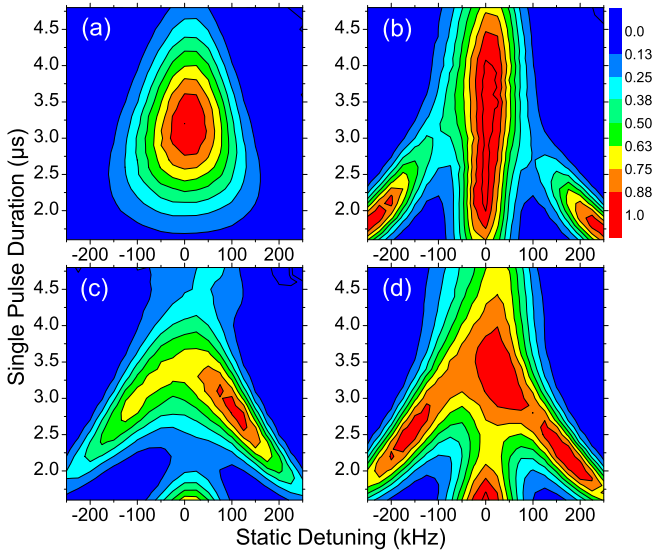}
\caption{(Color online) Experimentally measured rephasing efficiency of stored light vs. static detuning and single pulse duration for different rephasing sequences.
The frames use the same pulse sequences as in the theoretical simulations in Fig.~\ref{Fig-cp3}.
The Rabi frequency is kept fixed at $\Omega_{\text{RF}} \approx 2\pi \times 155\,\text{kHz}$.
Hence, a single $\pi$-pulse has a duration of $3.2\,\mu$s.
The slight asymmetry in the figures is due to some misalignment of a RF impedance matching circuit in the experiment.
}
\label{fig:exp_fig1}
\end{figure}

\begin{figure}
\includegraphics[width=1\columnwidth]{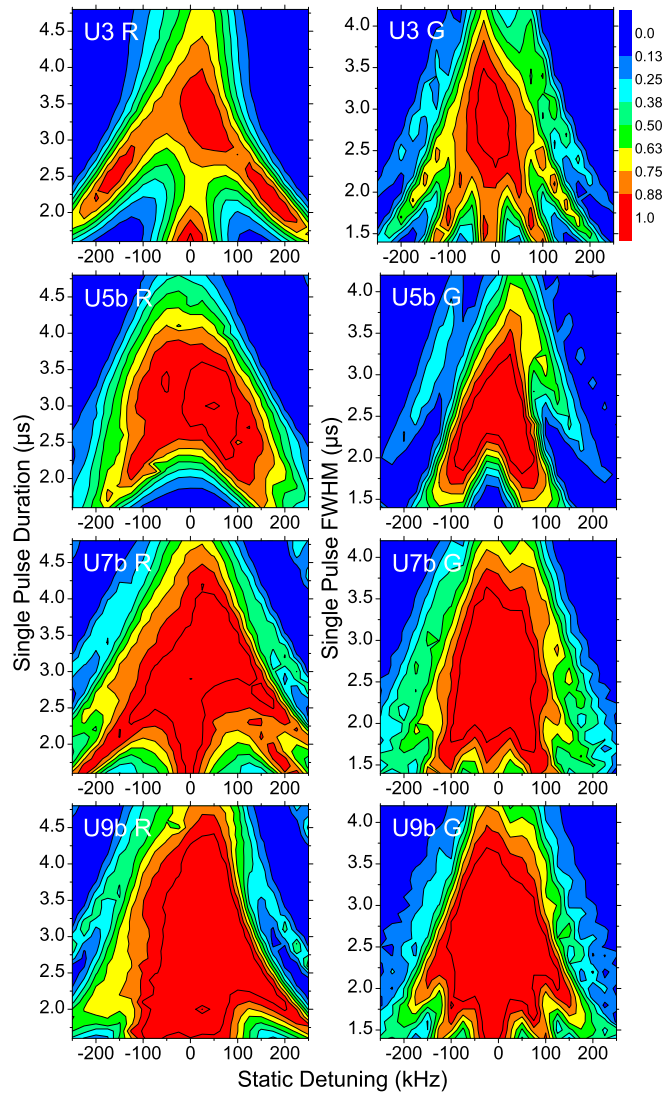}
\caption{(Color online)
Experimentally measured rephasing efficiency of stored light vs. static detuning and single pulse duration for different rephasing universal CP sequences from Table \ref{table:coef} (labeled in each frame).
The left column is for sequences of rectangular pulses and the right column is for sequences of Gaussian pulses.
For rectangular pulses, the pulse duration of a perfect $\pi$-pulse is 3.2 $\mu$s. This corresponds to a fixed Rabi frequency of $\Omega_{\text{RF}} \approx 2\pi \times 155\,\text{kHz}$.
For Gaussian pulses, the pulse duration (FWHM) of a perfect $\pi$-pulse is 2.8 $\mu$s. The corresponding peak Rabi frequency is also $\Omega_{\text{RF}} \approx 2\pi \times 155\,\text{kHz}$.} 
\label{fig:exp_fig2}
\end{figure}

Figure \ref{fig:exp_fig1} shows the experimentally measured rephasing efficiency for a single rephasing $\pi$-pulse and different CP rephasing sequences.
It uses the same pulse sequences as in the numerical simulations in Fig.~\ref{Fig-cp3}.
The experimental data agree remarkably well with the simulations in Fig.~\ref{Fig-cp3} and fully confirm the pronounced robustness of the universal CP sequence U3.

Next, we experimentally compare the behavior of universal CP sequences of different orders $n$=3, 5, 7, 9 in Fig. ~\ref{fig:exp_fig2}.
We apply CPs with rectangular and Gaussian temporal shape (see left and right columns in Figure ~\ref{fig:exp_fig2}).
The data for the universal CP sequence U3 R, already shown in Fig.~\ref{fig:exp_fig1} (d), serve as a benchmark to compare the CPs of different order.
As the data clearly show, the robustness of all universal CP sequences with regard to variations in the experimental parameters is much larger compared to a single $\pi$-pulse (compare Fig.~\ref{fig:exp_fig1} (a)).
As expected, the region of large rephasing efficiency increases with the number of pulses in the universal CP sequence.
The behavior is the same for rectangular and Gaussian pulses, i.e. the optimal phases in the CP sequence do not depend upon the pulse shape.
The experimental data clearly confirm the theoretical predictions:
 indeed, the overall shapes of the high-efficiency regions in Fig. ~\ref{fig:exp_fig2} look remarkably similar to the theoretical plots in Figs.~\ref{Fig-cp3} and \ref{Fig-all}. The ``duck-foot''-like pattern for U3, the ``boomerang-like'' pattern for U5b, and the ``pine-tree-like'' pattern for U7b are very nicely reproduced by the data.

In conclusion, we theoretically developed and experimentally demonstrated universal broadband CP sequences for robust high-fidelity population inversion.
These CPs compensate  simultaneous variations in \emph{all} experimental parameters (e.g. intensity or Rabi frequency, pulse duration, static detuning, Stark shifts, frequency chirp, etc.) and are applicable with \emph{any} arbitrary pulse shape.
The only theoretical assumptions are that the system has two states, the evolution is coherent and the single pulses in the sequence are identical. We experimentally confirmed the theoretical predictions by rephasing atomic coherences in a \pryso~crystal. In particular, our data demonstrate robust rephasing (hence, inversion) in a broad range of experimental parameters and for different pulse shapes. The efficiency and operation bandwidth of universal CPs are significantly larger compared to conventional $\pi$-pulses or non-universal CPs. As expected, the robustness increases for longer CP sequences. These universal CPs will be a highly accurate and robust tool for control of quantum systems, particularly valuable in situations with experimental uncertainties in the driving field and the environment.

 This work is supported by the Deutsche Forschungsgemeinschaft, the Alexander von Humboldt foundation, and the European Union's Seventh Framework Programme under REA Grant No. 287252.


\end{document}